\documentclass[11pt, a4paper]{article}
\pdfoutput=1
\usepackage{jheppub}
\usepackage{graphicx}
\usepackage{epsfig,amsmath,amsthm}
\usepackage{epstopdf}

\newcommand{\be}{\begin{equation}}
\newcommand{\ee}{\end{equation}}
\newcommand{\bea}{\begin{eqnarray}}
\newcommand{\eea}{\end{eqnarray}}
\def\bse{\begin{subequations}}
\def\ese{\end{subequations}}

\def\IZ{\relax\ifmmode\hbox{Z\kern-.4em Z}\else{Z\kern-.4em Z}\fi}
\def\half{\frac{1}{2}} \def\quart{\frac{1}{4}}

\def\del{{\partial}}

\def\lam{\lambda}

\def\presub{\vspace{.5cm} \noindent}

\def\bi{\begin{itemize}} \def\ei{\end{itemize}}

\def\({\left(} \def\){\right)}
\def\[{\left[} \def\]{\right]}
\def\<{\left<} \def\>{\right>}

\usepackage[OT2,T1]{fontenc}
\DeclareSymbolFont{cyrletters}{OT2}{wncyr}{m}{n}
\DeclareMathSymbol{\Sha}{\mathalpha}{cyrletters}{"58}
\usepackage{shuffle}

\title{1-loop Color structures and sunny diagrams}

\author{Barak Kol and Ruth Shir \\
{\it Racah Institute of Physics, Hebrew University, Jerusalem 91904, Israel} \\
{\tt barak.kol,ruth.shir@mail.huji.ac.il}
}

\abstract{The space of tree level color structures for gluon scattering was determined recently along with its transformation properties under permutations. Here we generalize the discussion to loops, demonstrating a reduction of an arbitrary color diagram to its vacuum skeleton plus rays. For 1-loop there are no residual relations and we determine the space of color structures both diagrammatically and algebraically in terms of certain sunny diagrams. We present the generating function for the characteristic polynomials and a list of irreducible representations for $3 \le n \le 9$ external legs. Finally we present a new proof for the 1-loop shuffle relations based on the cyclic shuffle and split operations.} 

\begin{document}
\maketitle

\section{Introduction}

Scattering amplitudes for QCD, or more general gauge theories, involve the color gauge group factors and kinematical data of the scattered particles. Therefore the first step in calculating a scattering amplitude of perturbative gauge theory is the separation of color from kinematics, written schematically as,
\be
 A_{tot} = \sum c_J\, A_J ~, 
  \label{Atot-scheme}
  \ee
where $A_{tot}$ represents the total amplitude, $c_J$ are the color structure coefficients which depend only on the color of the scattered particles and $J$ runs over all possible color structures, with $A_J$ the sub (or partial) amplitudes that depend only on the kinematical data (momenta and polarization). 
See the recent reviews \cite{DixonRev2011,DixonRev2013,ElvangHuang2013}. 

Investigating the symmetries of the space of color structure coefficients $c_J$ should give us information about the symmetries of the sub-amplitudes. These symmetries can be obtained from algebraic color identities not involving any kinematical data.

We consider scattering processes in theories where all particles are in the adjoint such as pure Yang-Mills or its supersymmetric generalizations. It was known for some time that the dimension of the space of color structures for the tree level scattering of $n$ gluons is $(n-2)!$ \cite{Cvitanovic-etal1981,DelDucaDixon-etal1999a,DelDucaDixon-etal1999b}. This was obtained using identities of the color structures and implies that there are at most $(n-2)!$ independent sub-amplitudes. In \cite{BCJ} it was shown that there are actually $(n-3)!$ independent sub-amplitudes once relations which involve kinematical factors are taken into account.  
The space of tree level color structures which we denote by $TCS_n$,\footnote{$TCS_n$ stands for Tree level Color Structures.} transforms under
$S_n$, the permutation group, which permutes the external gluons, yet only recently was this representation determined \cite{color}. To do that it was important to use color diagrams, made from oriented trivalent vertices, and the associated $f$-expressions \cite{Cvitanovic76,Cvitanovic-etal1981,DelDucaDixon-etal1999a,DelDucaDixon-etal1999b} which are built of the gauge group structure constants $f^{abc}$, in addition to the more popular trace-based color structures (which define the color ordered sub-amplitudes) \cite{BerendsGiele1987a,ManganoParkeXu1987,Mangano1988,ChanPaton1969}. The color diagrams, in turn, were recognized to be equivalent to the cyclic Lie operad, an object designed to study expressions composed of Lie generators, and whose representation spaces were already available in the mathematics literature \cite{GetzlerKapranov94}. The $S_4$ irreducible representation labelled by the partition $4=2+2$ was observed to be a recurring theme in scattering amplitudes \cite{2plus2}.
 
In this paper we generalize \cite{color} to include loops. We denote the space of 1-loop color structures with $n$ external gluons by $LCS_n$, 2-loop by $L^2CS_n$ and so on. By definition $L^0 CS_n \equiv TCS_n$. 1-loop color structures were studied in \cite{BernKosower90} in terms of traces and double traces and \cite{BDDK94} showed that the sub-amplitudes associated with double traces can be all expressed in terms of single trace sub-amplitudes, through certain relations which generalize the Kleiss-Kuijf relations \cite{KleissKuijf1988} from tree level to 1-loop (see \cite{BernDixonKosower1996} for a review). We shall refer to all such relations as shuffle relations, independent of the number of loops (the shuffle is also known as an ordered permutation). \cite{Naculich11} generalized  these relations to all loop orders, and \cite{EdisonNaculich12} obtained explicit results for $n \le 6$. A more comprehensive background on scattering amplitudes can be found in \cite{color} and the above-mentioned reviews.

In section \ref{sec:sunny} we determine the space $LCS_n$, in section \ref{sec:shuffle} we present a new proof for the 1-loop shuffle relations, and we summarize our results and conclude in section \ref{sec:summary}.

\section{Sunny diagrams}
\label{sec:sunny}
Color ordered sub-amplitudes transform in an interesting manner under permutations. In particular they must satisfy the Kleiss-Kuijf relations and their higher loop generalizations. Quite generally in the presence of symmetry it is standard practice to decompose the quantity under study into a part which is responsible for satisfying the permutation symmetry and a residual, invariant part which should be determined by other considerations. In our case this decomposition would schematically be 
\be
A_\sigma = S_\sigma \, A_0
\ee
where $\sigma \in S_n/\IZ_n$ is a cyclic ordering,  $S_\sigma$ is the permutation dependent factor (for instance the Parke-Taylor denominator) and $A_0$ is the permutation invariant part. From (\ref{Atot-scheme}) we see that the amplitudes transform in the same permutation representation labelled by $J$ as the color structures. Hence we set out to determine the representation of the color structures, and thereby determine the symmetry properties of the sub-amplitudes. It is true that at present various expressions for the sub-amplitudes $A$ are known and they were all found without assuming the symmetry, but rather they were shown to satisfy the symmetry \textit{a posteriori}. Still, the procedure outlined here is natural and may add insight to the issue.

\presub {\bf Reduction to ray diagrams}. The space of color structures can be identified with the space of color diagrams up to Jacobi relations \cite{Cvitanovic-etal1981}, where we use the same definitions and conventions as in \cite{color}. These diagrams apply to any theory where all the fields are in the adjoint representation of the gauge group including pure Yang-Mills (YM), pure supersymmetric YM (SYM) and ${\cal N}=4$ SYM. 

\begin{figure}[t!]
\centering \noindent
\includegraphics[width=8cm]{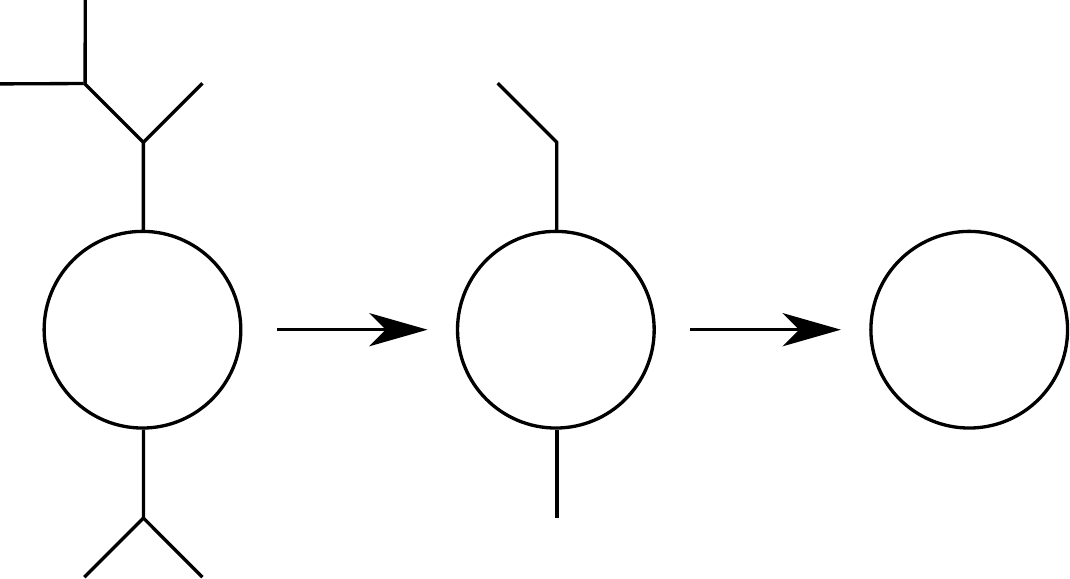}
\caption[]{An example of the pruning through repeated trimming of external legs, or leaves, to arrive at the vacuum skeleton sub-diagram.}
 \label{fig:trimming}
\end{figure}

To each color diagram we can associate a sub-diagram with no external legs, namely its vacuum skeleton. This can be done by an iterative pruning process where at each step all external legs, or leaves, are trimmed until none are left, at which point the pruning is complete. An example of this process is illustrated in fig. \ref{fig:trimming}. This pruning does not change the topology of the diagram, and the skeleton is non-empty if and only if the diagram has loops. 

Diagrammatically the Jacobi relations can be represented by the following move \be
\parbox{25mm}{\includegraphics[scale=1]{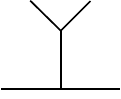}} = \parbox{25mm}{\includegraphics[scale=1]{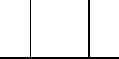}} - \parbox{25mm}{\includegraphics[scale=1]{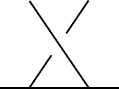}} ~.
\label{Jacobi}
\ee

\begin{figure}[t!]
\centering \noindent
\includegraphics[width=10cm]{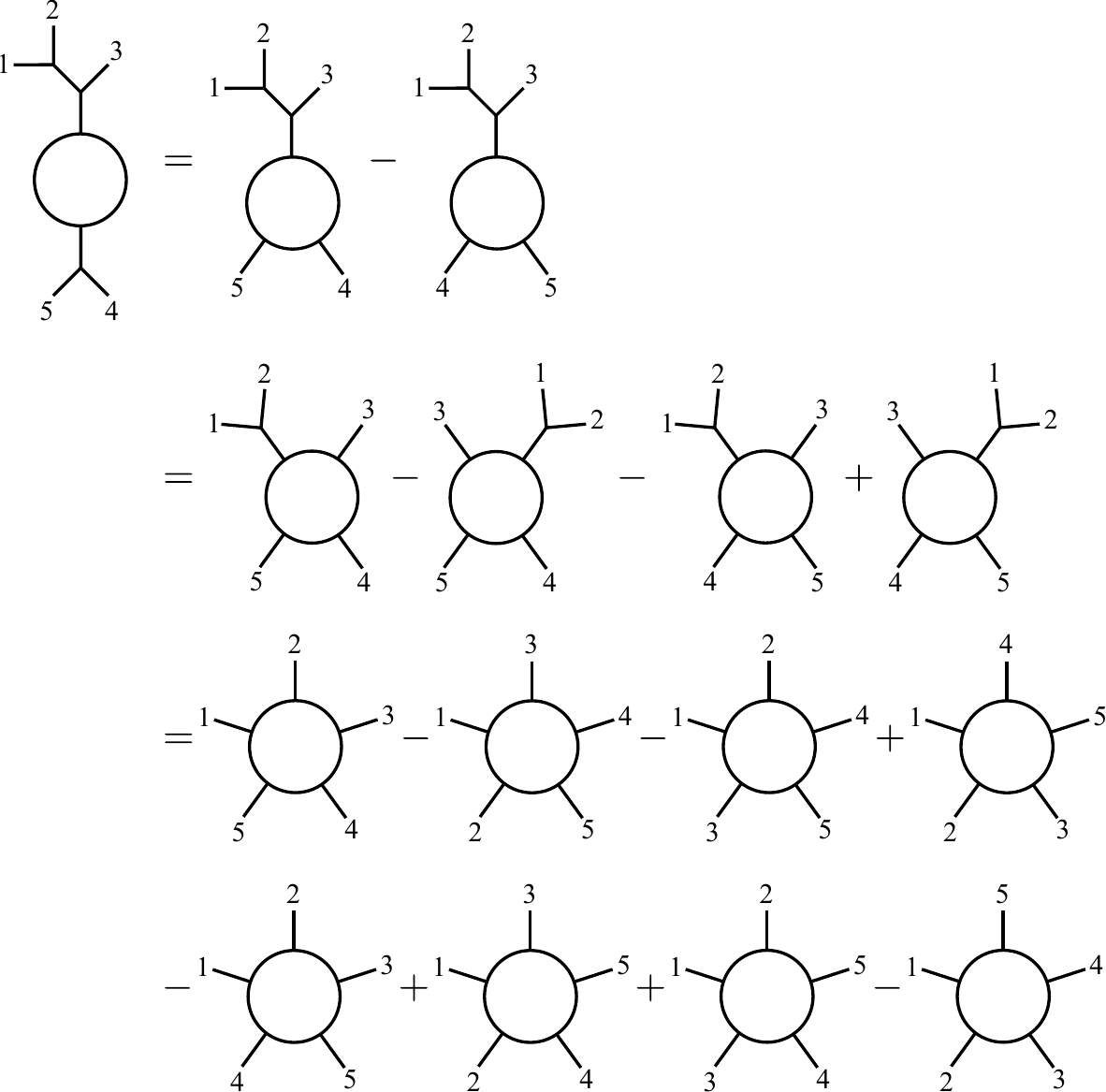}
\caption[]{An example for the outcome of Jacobi planting operating on a 1-loop color diagram with $n=5$ external legs, converting it to a linear combination of sunny diagrams (1-loop ray diagrams).}
\label{fig:Jacobi-planting}
\end{figure}

We proceed to reduce a general color diagram through a process of ``Jacobi planting'' as follows. 
 Once the vacuum skeleton is found one identifies edges which end on it, but are not part of it, namely edges which are directly attached to the vacuum skeleton, and one operates on them with the Jacobi moves. Repeating this process the diagram can be converted to a linear combination of diagrams where all trivalent vertices are within the vacuum skeleton. We shall refer to such diagrams as ray diagrams and they  have the property that the pruning process described above would end after a single step. In other words we define a ray diagram as a diagram where all external legs are attached directly to the vacuum skeleton. See fig. \ref{fig:Jacobi-planting} for an example illustrating the outcome of this process. In summary, Jacobi planting is a process that converts any color diagram to a linear combination of ray diagrams all with the same vacuum skeleton as the original diagram. Hence $L^\ell CS_n$ can be realized diagrammatically by the space of $\ell$-loop ray diagrams up to residual Jacobi identities -- ones which preserve the ray nature of the diagram.

\presub {\bf Sunny diagrams}. For 1-loop diagrams the vacuum skeleton is unique -- it is the circle. We call the ray diagrams over it ``sunny diagrams''. For example, fig. \ref{fig:Jacobi-planting} depicts sunny diagrams with $n=5$ as an outcome of Jacobi planting over a more general 1-loop diagram. For $n>3$ , any Jacobi move would spoil the ray nature and hence there are no residual Jacobi relations; we find that \be
 LCS_n = \{ \mbox{space of sunny diagrams} \} ~.
 \label{lcsn-diag}
 \ee

The cases $n=2,3$ are special. For $n=2$ we have \be
\parbox{32mm}{\includegraphics[scale=0.5]{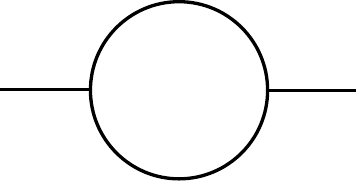}}  = C_2\, \parbox{32mm}{\includegraphics[scale=0.5]{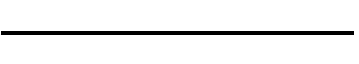}} ~,
 \label{n2loop}
 \ee
 where $C_2$ is the second Casimir of the adjoint representation, and the proportionality constant depends on the normalization chosen for the structure constants which we take as in \cite{PeskinSchroeder}, where (\ref{n2loop}) is equivalent to their eq. (15.93) which reads $f^{a c d}\, f^{b c d }=C_2\, \delta^{a b}$.

For $n=3$, application of the Jacobi relation to any one of the edges, together with (\ref{n2loop}) gives \be
\parbox{25mm}{\includegraphics[scale=0.5]{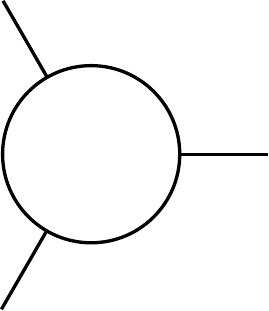}} = \frac{C_2}{2}\, \parbox{25mm}{\includegraphics[scale=0.5]{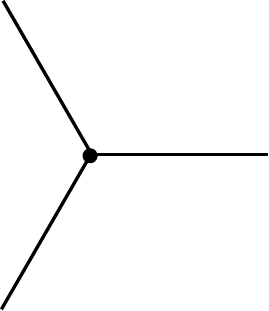}} ~.
 \label{n3loop}
 \ee
We note that (\ref{n2loop},\ref{n3loop}) hold for any group, not necessarily $SU(N_c)$.

\presub {\bf Algebraic determination of $LCS_n$}. Each sunny diagram encodes an $f$-expression. Clearly the expression is unchanged by a cyclic permutation of the external legs. In addition a reflection of the whole diagram inverts all orientations and changes the sign by $(-)^n$. The rotations and reflections together consist of the dihedral group, and the above described sign defines a representation of the dihedral group which we shall call the sign representation.  By definition the symmetries imply that $LCS_n$ can be described algebraically by an induced representation \be
LCS_n = {\rm Ind}_{D_n}^{S_n}\, {\rm sgn} ~~,
\label{lcsn-algeb}
\ee
 where $D_n$ denotes the dihedral group, and ${\rm sgn}$ is its above-mentioned sign representation. Eq. (\ref{lcsn-algeb}) implies that \be
 {\rm dim}(LCS_n)= \frac{|S_n|}{|D_n|} = \half \, (n-1)! ~.
 \label{dim-lcsn}
\ee

We note that single traces have precisely the same symmetries as the sunny diagrams. As $TCS_n$ can be identified with a subspace of the single traces we see that \be
 TCS_n \subset LCS_n ~.
 \label{tcsn-in-lcsn}
 \ee
This is consistent with the expectation that sub-amplitudes generated at tree level could receive loop corrections.

\presub {\bf The characteristic polynomial}. Using the algebraic identification of $LCS_n$ as an $S_n$ representation (\ref{lcsn-algeb}), we can proceed to find its character, or more conveniently its characteristic polynomial. The character fully describes the symmetry properties of the color structures under permutations, and hence also the symmetry properties of the corresponding sub-amplitudes. In particular, it allows us to decompose spaces of color structures into their irreducible components. The derivation is described in appendix \ref{app:ch}. The end result is that the generating function of characteristic polynomials is given by \be
ch(LCS) \equiv \sum_n  ch(LCS_n) = \half \sum_{n=1}^{\infty} \frac{\varphi(n)}{n}\, \log(1-p_n) +  \frac{(1-p_1)^2}{4(1-p_2)} ~, 
\label{gen-func}
\ee
 where $\varphi(n)$ is Euler's totient function whose value is the number of $k$s within the range  $1 \le k \le n$ which are relatively prime to $n$, see for example  \cite{wiki-totient}, and $p_n=\sum_i \(x_i\)^n$ are symmetric power sums in some $x$ variables as in \cite{color} subsection 3.2. By expanding the generating function to a power series in $p_1, p_2, p_3, \dots$ and assigning a weight $l$ to each $p_l$ one can read off the characteristic polynomials for each $n$. For $n$ odd $n=2 m+1$ \be
 ch(LCS_n) = \frac{1}{2 n} \sum_{d|n} \varphi(d)\, p_d^{n/d} - \half \, p_1\, p_2 ^m ~,
 \label{ch-odd} \ee
 while for $n$ even $n= 2 m$ we have \be
 ch(LCS_n) = \frac{1}{2 n} \sum_{d|n} \varphi(d)\, p_d^{n/d} + \quart \, \( p_1^2\, p_2^{m-1} + p_2 ^m \)  ~.
 \label{ch-even} \ee

Given the characteristic polynomial, we can decompose the representation into its irreducible components (irreps). First the associated character is computed through \be
\chi(\lam) = \( 1 \del_{p_1}\)^{\lam_1} \(2\, \del_{p_2}\)^{\lam_2} \dots ch(p_1,p_2,\dots) |_{0=p_1=p_2=\dots} ~,
\label{chi}
\ee
 where $\lam=(1^{\lam_1}\, 2^{\lam_2}\, \dots)$ is any cycle structure type in $S_n$, namely a partition which labels any $S_n$ conjugacy class.
Next the representation is decomposed using the character tables and computerized calculation \cite{GAP4}. The results for $3 \le n \le 9$ are presented in table \ref{tab:irrep},\footnote{We thank S. Naculich for correcting a typo which appeared in a previous version of this table.} where due to (\ref{tcsn-in-lcsn}) it suffices to list the representation added \be
 \delta LCS_n := LCS_n - TCS_n
 \ee
 rather than all of $LCS_n$. 
 These results were tested by confirming that the dimensions add up to (\ref{dim-lcsn}), by confirming that the components of $TCS_n$ are included as implied by (\ref{tcsn-in-lcsn}), and by comparison with existing results for $4 \le n \le 6$  \cite{EdisonNaculich12}.

In \cite{color} we noticed that for some $n$, $TCS_n$ is self-dual under Young conjugation. By Young conjugation we mean the interchange of rows and columns in the Young diagrams of the irrep components, or equivalently tensoring the representation with the sign representation of $S_n$. We now observe that for $n=5,9$ not only is $TCS_n$ self-dual \cite{color} but so is $\delta LCS_n$ (and hence also $LCS_n$). This may hint at a pattern where for some $n$ the color structures are self-dual. The significance of this observation is not clear at present.

\begin{table}[t!]
\caption{Multiplicities of the irreducible representation of $S_n$ for tree level color structures $TCS_n$ \cite{color}, repeated here for reference, and $\delta LCS_n := LCS_n - TCS_n$ where $LCS_n$ are 1-loop color structures. The notation $\lambda=a^i b^j...$ describes a Young diagram with $a$ boxes in each of the first $i$ rows and $b$ boxes in each of the next $j$ rows, etc. or equivalently a partition of $n$, $i \cdot a + j \cdot b + \dots = n$.}
\centering
\begin{tabular}{l  p{7cm}  p{7cm} }
\hline
$n$ & $TCS_n$ & $\delta LCS_n$\\
\hline
3 & $(1^3)$ & 0\\
\hline
4 & $(2^2)$ & (4)\\
\hline
5 & $(31^2)$ & $(31^2)$\\
\hline
6 & $(2^3)+(1^33)+(24)$ & $(1^42)+(2^3)+(123)+(24)+(6)$\\
\hline
7 & $(1^32^2)+(13^2)+(124)+(1^25)$ & $(1^7)+(1^32^2)+(12^3)+3(1^223)+2(13^2)+(1^34)+(124)+ (34) + 2(1^25)$\\
\hline
8 & $(1^42^2)+(2^4)+(1^323)+(12^23)+2(1^23^2)+(1^44)+(1^224)+2(2^24)+(134)+(4^2)+(1^35)+(125)+(26)$ & $(1^42^2)+2(2^4)+(1^53)+3(1^323)+3(12^23)+3(1^23^2)+(23^2)+3(1^44)+3(1^224)+5(2^24)+2(134)+2(4^2)+(1^35)+3(125)+(35)+2(26)+(8)$\\
\hline
9 & $(1^32^3)+(1^63)+(1^423)+3(1^22^23)+(2^33)+(1^33^2)+2(123^2)+2(3^3)+(1^54)+3(1^324)+3(12^24)+3(1^234)+2(234)+(14^2)+3(1^225)+(2^25)+3(135)+(1^36)+(126)+(36)+(1^27)$ & $2(1^32^3)+(12^4)+3(1^63)+4(1^423)+9(1^22^23)+2(2^33)+2(1^33^2)+7(123^2)+4(3^3)+2(1^54)+9(1^324)+9(12^24)+9(1^234)+7(234)+2(14^2)+(1^45)+9(1^225)+2(2^25)+9(135)+(45)+2(1^36)+4(126)+2(36)+3(1^27)$ \\ 
\end{tabular}
\label{tab:irrep} 
\end{table}

\section{Shuffle relations}
\label{sec:shuffle} 

The authors of ref. \cite{BDDK94} discovered and proved that for $SU(N_c)$ theories at 1-loop the double trace sub-amplitudes can be expressed in terms of the single trace sub-amplitudes. This is consistent with our realization that the space $LCS_n$ is equivalent to the space of single traces (\ref{lcsn-algeb},\ref{tcsn-in-lcsn}). In this section we present a new proof for this relation based on the cyclic shuffle (also known as cyclic ordered permutation or COP) and split operations, to be defined below, in analogy with the use of the ordinary shuffle and split operations to prove the tree level Kleiss-Kuijf relations in \cite{color}, subsection 2.3. 

As in \cite{color} we proceed by relating the $f$ and trace decompositions. The total amplitude at any loop level given schematically in (\ref{Atot-scheme}) is a linear combination of the sub-amplitudes $A_J$ with color structure coefficients $c_J$. There are two known ways to expand the total amplitude in terms of the color structure coefficients: one is using the $f$-basis where each color coefficient $c^f$ is constructed out of the group structure constants $f^{abc}$, and the other is using the trace basis where the color coefficients $c^t$ are constructed from traces of $T^aT^b...$ where the $T$'s are the generators of the gauge group at a given representation. For each of the sets of color coefficients there is a corresponding set of sub-amplitudes which we denote $A^f$ and $A^t$ respectively. We will see that relations between the two forms of color coefficients give us relations between the sub-amplitudes.

Once we know the form of the color coefficient in the $f$-basis we can transform to the trace basis using the relation 
\be
 i\, f^{a b c} \to {\rm Tr}\(T^a T^b T^c - T^c T^b T^a\) 
 \label{f-to-tr}
 \ee 
for each $f^{abc}$ in $c^f$.
It is useful to represent this relation graphically by 
\be
 \parbox{70mm}{\includegraphics[scale=0.5]{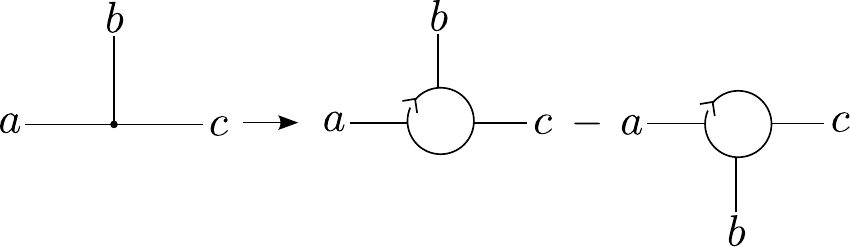}}~~~~.
\label{graph-f-to-tr}
\ee
We then perform the summation over internal lines using the $SU(N_c)$ identity 
\be
\( T^a \)_i^j \( T^a \)_k^l = \delta_i^l \delta_k^j - \frac{1}{N_c}\delta_i^j \delta_k^l ~.
\label{SUn-completeness}
\ee
When all fields are in the adjoint representation the $\frac{1}{N_c}$ term can be omitted such that the relation can be represented graphically as
\be
\parbox{40mm}{\includegraphics[scale=0.8]{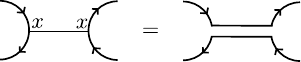}} ~.
\label{graph-completeness}
\ee

For one loop amplitudes we have seen that the basis of $c^f$ color coefficients are the sunny diagrams (\ref{lcsn-diag}), namely
\be
c^f(12...n)\equiv \raisebox{-1.5em}{\includegraphics[scale=0.3]{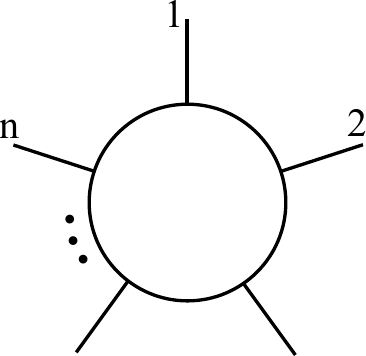}} \equiv f^{x_n 1 x_1}f^{x_1 2 x_2}...f^{x_{n-1} n x_n} ~.
\ee
We will use (\ref{graph-f-to-tr}) and (\ref{graph-completeness}) to transform each $c^f$ of this form into a sum of trace color coefficients $c^t$.

To recognize the permutations which appear in the sum, we will need some mathematical concepts which we now turn to define. A cyclic word is a word defined modulo shifts, and we shall mostly write it within parentheses. Hence for example $(abc) \equiv (bca)$. The cyclic split, $\delta_c$, of a cyclic word $\sigma$ is a sum over all pairs of cyclic words $(\alpha,\beta)$ such that each letter of $\sigma$ is either in $\alpha$ or in $\beta$ and each preserves the original cyclic ordering of the letters in $\sigma$.
For example:
\bea
\delta_c\( (abcd) \)&=&\( (abcd), \emptyset \)+\( (a), (bcd) \)+\( (b), (acd) \)+\( (c), (abd) \)+\( (d), (abc) \) + \\
&+&\(\emptyset, (abcd)\)+\( (bcd), (a) \)+ \( (acd), (b) \)+\( (abd), (c) \) + \( (abc), (d)\) + \nonumber \\
&+& \( (ab), (cd) \)+\( (ac) ,(bd) \)+\((ad), (bc)\) + \( (cd), (ab) \)+\( (bd),(ac)\)+\( (bc),(ad) \) \nonumber
\eea
Just like the ordinary split $\delta$, the number of terms in $\delta_c(\sigma)$ is $2^{|\sigma|}$, where $|\sigma|$ is the length of $\sigma$ namely the number of its letters. We continue to call this operator ``split'' as in \cite{color} because it splits a word into pairs. In the mathematics literature it is also known as the co-shuffle or the unshuffle. The subscript $c$ reminds us that this is the cyclic split operation; it can be omitted if clear from context. 

The cyclic shuffle, $\shuffle_c$, \footnote{
The cyclic shuffle has appeared already in the physics literature \cite{ReuschleWeinzierl13} where it was denoted by a circle within a circle, but we use a notation which stresses the analogy with the non-cyclic shuffle.} 
of two cyclic words $\alpha$ and $\beta$ is defined as the sum of all cyclic words $\sigma$ which can be written by distributing the word $\alpha$ through the word $\beta$ while keeping the cyclic ordering within each of them. 
For example
\bea
(abc) \shuffle_c (de) &=& (abcde) + (abced) + (abdce) + (abecd) + (abdec) + (abedc) \nonumber \\
 &+& (adbce) + (aebcd) + (adbec) + (aebdc) + (adebc) + (aedbc) \nonumber
\eea
The number of terms in $\alpha \shuffle_c \beta$ is $(|\alpha|+|\beta|-1)!/(|\alpha|-1)!/(|\beta|-1)!\,$. 
Just as the ordinary shuffle and split  $\shuffle$ and $\delta$ are adjoint, so are the cyclic operators; namely, \be
 \delta_c(\sigma) \cdot (\alpha,\beta) = \sigma \cdot ( \alpha \shuffle_c \beta) ~,
 \label{adjoint}
 \ee
where all the words are cyclic and the inner product is defined in comment 7 of \cite{color}. This property means that a cyclic word $\sigma$ will appear in the cyclic shuffle of two cyclic words $\alpha$ and $\beta$ if and only if $\alpha$ and $\beta$ appear in the cyclic split of $\sigma$.

After these preparations we turn to the proof which is rather short. 

To transform from the $f$ basis to the trace basis we perform the following steps, represented schematically by 
\bea
c^f(\sigma) &=&\raisebox{-2.5em}{\includegraphics[scale=0.35]{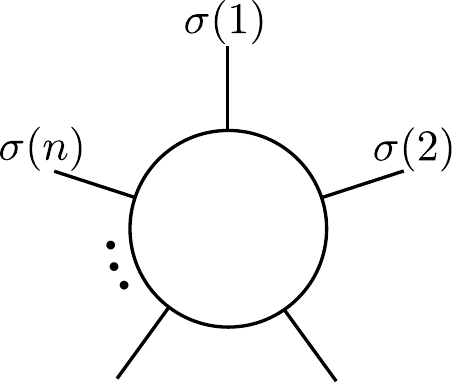}}  \to \sum_{(\alpha,\beta)\in \delta_c(\sigma)}(-)^{|\beta|}\raisebox{-2.5em}{\includegraphics[scale=0.35]{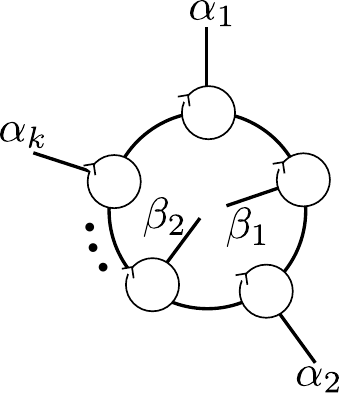}} \\
&\to& \sum_{(\alpha,\beta)\in \delta_c(\sigma)}(-)^{|\beta|}\raisebox{-2.5em}{\includegraphics[scale=0.35]{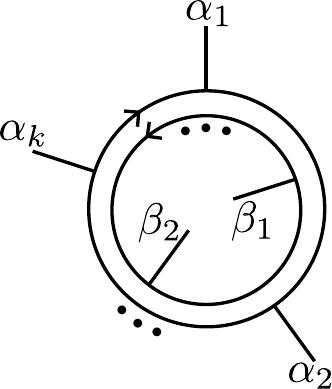}} =\sum_{(\alpha,\beta)\in \delta_c(\sigma)}(-)^{|\beta|}{\rm Tr}(\alpha)\,{\rm Tr}(\beta^T)
\eea
where in the first step we used (\ref{graph-f-to-tr}) to get a sum over all the ways to split the cyclic word $\sigma$ into two cyclic words: $\alpha$ (pointing outwards) and $\beta$ (pointing inwards). Note that each letter in $\beta$ contributes a minus sign according to (\ref{graph-f-to-tr}). In the second step we used (\ref{graph-completeness}) to perform the sum over internal lines and get a double lined circle which represents a double trace ${\rm Tr}(\alpha)\,{\rm Tr}(\beta^T)$. Notice that we read the order of the generators in each trace according to the arrow on the line and therefore the cyclic word $\beta$ is read in the opposite cyclic order $\beta^T=(\beta_n \beta_{(n-1)}...\beta_2 \beta_1)$.

Now we have
\bea
A_{tot}&=&\sum_{\sigma}c^f(\sigma)A^f(\sigma)=\sum_{\sigma}\Big[\sum_{(\alpha,\beta)\in \delta_c(\sigma)}(-)^{|\beta|}{\rm Tr}(\alpha)\,{\rm Tr}(\beta^T) \Big] A^f(\sigma)\\
&=&\sum_{(\alpha,\beta)}\Big[\sum_{\sigma\in\alpha\shuffle_c\beta}(-)^{|\beta|}A^f(\sigma)\Big]{\rm Tr}(\alpha)\,{\rm Tr}(\beta^T) ~.
\eea
 In the third equality we used (\ref{adjoint}) according to which two cyclic words $\alpha$ and $\beta$ appear in the split of a cyclic word $\sigma$ if and only if $\sigma$ appears in the corresponding shuffle.

We recognize that by definition the term which appears in the square brackets is the sub-amplitude in the trace basis with the double trace color coefficient ${\rm Tr}(\alpha)\,{\rm Tr}(\beta^T)$. Hence we identify
\be
A^t({\alpha},{\beta^T})=\sum_{\sigma\in\alpha\shuffle_c\beta}(-)^{|\beta|}A^f(\sigma) ~.
\label{At-from-Af}
\ee
In the case $\beta=\emptyset$ we find 
\be
A^t(\sigma)=A^f(\sigma)
\ee
which we substitute back into (\ref{At-from-Af}) to obtain
\be
A^t({\alpha},{\beta})=\sum_{\sigma \in \alpha \shuffle_c \beta^T} (-)^{|\beta|} A^t(\sigma) ~.
\ee
 These are the sought after shuffle relations of \cite{BDDK94} thereby completing the proof.
 
 \presub {\bf Comparison with existing proofs}. The proof in \cite{BDDK94} is given in two approaches -- one is motivated by open strings and the other is intrinsically field theoretic. In each approach certain classes of excluded terms are identified, as shown in fig. 8 and 9 there, and their total contribution is shown to cancel. In addition a certain $U(P) \times U(N_c - P)$ decoupling is discussed. In the current proof these notions are mostly absent and unnecessary.
 
 The method of \cite{ReuschleWeinzierl13} is based on two categories of operations: double-ring structures in one-loop color-flow diagrams and $U(1)$-gluons. Again these notions appear to be absent in the current proof. In addition while the cyclic shuffle plays a central role in both approaches, we could not identify the use of the split operation in them.
  
\section{Summary}
\label{sec:summary}

This paper generalizes \cite{color} to loops and applies to any theory where all the fields are in the adjoint representation. For a general number of loops we were able to reduce the space of color structures to the space of ``vacuum skeletons'' with rays together with some residual relations inherited from the Jacobi relations, see fig. \ref{fig:Jacobi-planting}.

In the 1-loop case we fully determined  $LCS_n$, the space of loop color structures. As there are no residual relations $LCS_n$ is given by ``sunny diagrams'', such as the outcome of fig. \ref{fig:Jacobi-planting}. Such diagrams are labelled by cyclic orderings of the $n$ external legs, up to reversal, and hence ${\rm{dim}}(LCS_n) = (n-1)!/2$. Algebraically this is equivalent to an $S_n$ representation induced by the dihedral group (\ref{lcsn-algeb}), meaning that it is in the same representation as single traces. Since $TCS_n$ can be identified with a subspace of the single traces we see that $TCS_n \subset LCS_n$ (\ref{tcsn-in-lcsn}) as we might expect in order that all the sub-amplitudes which appear at tree level may receive 1-loop corrections. The generating function of the characteristic polynomial is given by (\ref{gen-func}), and the individual polynomials are given by (\ref{ch-odd},\ref{ch-even}). For $3 \le n \le 9$, the representations were decomposed into irreducible components through the use of the character table and the results are listed in table \ref{tab:irrep}.

At 1-loop the Kleiss-Kuijf relations generalize to the relations found in \cite{BDDK94}. We collectively refer to these as ``shuffle relations''. In section \ref{sec:shuffle} we used the notions of cyclic shuffle
 and split to present an elegant proof of the 1-loop shuffle relations based on \cite{color,DelDucaDixon-etal1999b}. 
  
Our main results are the generating function (\ref{gen-func}) and the list of irrep decompositions in table \ref{tab:irrep}. We tested them successfully as described below (\ref{chi}) and we believe that they are novel. These are rather basic quantities associated with scattering amplitudes, and therefore are of intrinsic interest. In addition they are interesting through their relation to the literature as reflected by the list of references. We believe these results appear here for the first time.
 
Higher loop color structures are left for future work. Also, seeking physical implications for the irrep decomposition remains an open question.

\subsection*{Acknowledgments}

This research was supported by Israel Science Foundation grant no. 812/11 and it is part of the Einstein Research Project "Gravitation and High Energy Physics", which is funded by the Einstein Foundation Berlin.

\appendix

\section{Derivation of characteristic polynomials}
\label{app:ch}

In this section we derive the characteristic polynomial, $ch$ for the induced representation (\ref{lcsn-algeb}).

As a first step we calculate $ch$ for the representation induced by the trivial representation of the cyclic group $\IZ_n$. For a general representation induced into $S_n$ the characteristic polynomial $ch$ is a bit simpler than the Frobenius formula for the character; it is given by, \be 
 ch\( \mbox{Ind}_H^{S_n}\, \xi \) = \frac{1}{|H|} \sum_{h \in H} \xi(h)\, p_{\lambda(h)}    ~,
\label{ch-induced}
\ee
where $\lambda(h)$ specifies the cycle structure of $h$, namely $\lambda = 1^{\lambda_1}\, 2^{\lambda_2} \dots$ is the partition of $n$ which labels the conjugacy class of $h$, and $p_{\lambda(h)}:= p_1^{\lambda_1}\, p_2^{\lambda_2} \dots$. From this one finds \be
ch \( \mbox{Ind}_{\IZ_n}^{S_n}\, 1 \) = \frac{1}{n} \sum_{j=0}^{n-1} p_{\lambda(r^j)} = \frac{1}{n} \sum_{d|n} \varphi(d)\, p_d^{n/d} ~,
\label{ch-cyclic}
\ee
where $r$ is the cyclic generator and $\varphi(n)$ is Euler's totient function, see below (\ref{gen-func}).

Next we proceed to the dihedral group. For $n$ odd, so $n=2 m+1$, all  reflections $\sigma$ have a single fixed point and hence their cycle structure is $\lambda(\sigma) =(1\, 2^m)$. In the sign representation, $\mbox{sgn}_{2m+1}(\sigma)=-1$ and $ch$ is given by (\ref{ch-odd}). For $n$ even, so $n= 2 m$, the reflections are divided between those with no fixed points and hence with cycle structure type $(2^m)$ and those with 2 fixed points and hence of type $(1^2\, 2^{m-1})$. For all of them, $\mbox{sgn}_{2m}(\sigma)=+1$ and $ch$ is given by (\ref{ch-even}). The odd and even expressions add elegantly  to the generating function (\ref{gen-func}).

\bibliographystyle{unsrt}

\end{document}